\documentclass[twocolumn,superscriptaddress,aps,preprintnumbers,amsmath,amssymb,prl,nofootinbib]{revtex4}

\usepackage{graphicx}
\usepackage{epstopdf}
\usepackage{dcolumn}% Align table columns on decimal point
\usepackage{bm}% bold math
\usepackage{hyperref}
\usepackage{color}

\begin{document}
%%%%%%%%%%%%%%%%%%%%%%%%%%%%%%%%%%%%%%%%%%%

\def\a{\alpha}
\def\b{\beta}
\def\c{\varepsilon}
\def\d{\delta}
\def\e{\epsilon}
\def\f{\phi}
\def\g{\gamma}
\def\h{\theta}
\def\k{\kappa}
\def\l{\lambda}
\def\m{\mu}
\def\n{\nu}
\def\p{\psi}
\def\q{\partial}
\def\r{\rho}
\def\s{\sigma}
\def\t{\tau}
\def\u{\upsilon}
\def\v{\varphi}
\def\w{\omega}
\def\x{\xi}
\def\y{\eta}
\def\z{\zeta}
\def\D{\Delta}
\def\G{\Gamma}
\def\H{\Theta}
\def\L{\Lambda}
\def\F{\Phi}
\def\P{\Psi}
\def\S{\Sigma}

\def\o{\over}
\def\beq{\begin{eqnarray}}
\def\eeq{\end{eqnarray}}
\newcommand{\gsim}{ \mathop{}_{\textstyle \sim}^{\textstyle >} }
\newcommand{\lsim}{ \mathop{}_{\textstyle \sim}^{\textstyle <} }
\newcommand{\vev}[1]{ \left\langle {#1} \right\rangle }
\newcommand{\bra}[1]{ \langle {#1} | }
\newcommand{\ket}[1]{ | {#1} \rangle }
\newcommand{\EV}{ {\rm eV} }
\newcommand{\KEV}{ {\rm keV} }
\newcommand{\MEV}{ {\rm MeV} }
\newcommand{\GEV}{ {\rm GeV} }
\newcommand{\TEV}{ {\rm TeV} }
\newcommand{\1}{\mbox{1}\hspace{-0.25em}\mbox{l}}
\def\diag{\mathop{\rm diag}\nolimits}
\def\Spin{\mathop{\rm Spin}}
\def\SO{\mathop{\rm SO}}
\def\O{\mathop{\rm O}}
\def\SU{\mathop{\rm SU}}
\def\U{\mathop{\rm U}}
\def\Sp{\mathop{\rm Sp}}
\def\SL{\mathop{\rm SL}}
\def\tr{\mathop{\rm tr}}

\def\IJMP{Int.~J.~Mod.~Phys. }
\def\MPL{Mod.~Phys.~Lett. }
\def\NP{Nucl.~Phys. }
\def\PL{Phys.~Lett. }
\def\PR{Phys.~Rev. }
\def\PRL{Phys.~Rev.~Lett. }
\def\PTP{Prog.~Theor.~Phys. }
\def\ZP{Z.~Phys. }

%%%%%%%%%%%%%%%%%%%%%%%%%%%%%%%%%%%%%%%%%%%%%%%%%%%%%%%%%%%%%%%

\title{
Chaotic Inflation with a Fractional Power-Law Potential\\
in Strongly Coupled Gauge Theories
}

\author{Keisuke Harigaya}
\affiliation{Kavli IPMU, TODIAS, University of Tokyo, Kashiwa, 277-8583, Japan}
\author{Masahiro Ibe}
\affiliation{Kavli IPMU, TODIAS, University of Tokyo, Kashiwa, 277-8583, Japan}
\affiliation{ICRR, University of Tokyo, Kashiwa, 277-8582, Japan}
\author{Kai Schmitz}
\affiliation{Kavli IPMU, TODIAS, University of Tokyo, Kashiwa, 277-8583, Japan}
\author{Tsutomu T.~Yanagida}
\affiliation{Kavli IPMU, TODIAS, University of Tokyo, Kashiwa, 277-8583, Japan}

\begin{abstract}
Models of chaotic inflation with a fractional power-law potential
are not only viable but also testable in the foreseeable future.
We show that such models can be realized 
in simple strongly coupled supersymmetric gauge theories.
In these models, the energy scale during inflation is dynamically generated by 
the dimensional transmutation due to the strong gauge dynamics.
Therefore, such models not only explain the origin of the fractional power 
in the inflationary potential but also provide a reason why the energy
scale of inflation is much smaller than the Planck scale.
\end{abstract}

\date{\today}
\maketitle
\preprint{IPMU 12-0215}
%\preprint{ICRR-report-xxx}

\section{Introduction}

Cosmic inflation\,\cite{Guth:1980zm} is a very successful paradigm 
of modern cosmology which explains the origin  of the anisotropies of
the Cosmic Microwave Background (CMB) 
as well as of the Large Scale Structure of the Universe\,\cite{Mukhanov:1981xt,Lyth:1998xn}.
At present, a realistic and complete theory of inflation is, however,
still pending and hence, inflationary model building remains one of
the most important tasks of particle physics and cosmology.
 
Among the various classes of inflation models proposed so far, 
the chaotic inflation scheme\,\cite{Linde:1983gd} is one of the most attractive classes
since it can realize an inflationary expansion 
even in the presence of large quantum fluctuations at the Planck time.
Moreover, the large field values typically encountered in models of chaotic inflation
imply a large contribution from gravitational waves to the CMB power
spectrum\,\cite{Starobinsky:1985ww}, rendering these models testable in
the foreseeable future.
However, according to the precise observations of the CMB anisotropies,
the simplest versions of chaotic inflation, i.e.\
the models with a quadratic potential or a quartic potential, 
are now  somewhat disfavored\,\cite{Komatsu:2010fb}.
With the forthcoming data provided by 
the Planck satellite experiment\,\cite{Planck:2006aa},
the constraints on those simplest versions will be improved upon.

In light of this situation, a more general version of chaotic inflation 
has been gathering attention, in which the inflaton potential comes with a fractional power.
As analyzed in Refs.\,\cite{Alabidi:2010sf}, the models of chaotic inflation
with a fractional power-law potential are more favored than 
the simplest versions of chaotic inflation.
Despite such successes, these models, however, lack a firm field theoretical foundation,
which apparently seems difficult to be achieved from regular field theories.%
\footnote{A recipe for embedding chaotic inflation with a fractional
power-law potential into supergravity is provided in Refs.\,\cite{Kallosh:2010ug,Kallosh:2010xz};
see also Ref.\,\cite{Takahashi:2010ky}, in which a fractional power-law potential 
is obtained from  a running kinetic term for the inflaton.
For fractional power-law potentials derived in string theories, see Ref.\,\cite{Silverstein:2008sg}.}

In this paper, we show that such fractional power-law chaotic inflation models can be realized 
in simple strongly coupled supersymmetric gauge theories.
There,  the energy scale during inflation is generated by 
the dimensional transmutation due to the strong gauge dynamics.
Thus, these models not only explain the origin of the fractional power 
in the inflationary potential but also provide a reason why the energy scale
of inflation is much smaller than the Planck scale.%
\footnote{For examples of models in which the scale of inflation
is generated dynamically, see Refs.\,\cite{Dimopoulos:1997fv}.}

The organization of the paper is as follows.
First, we derive the fractional power-law potential for the inflaton in
strongly coupled gauge theories.
Next, we discuss distortions of the inflaton potential due to
supergravity contributions.
Then, we outline the phenomenology of chaotic inflation with the dynamically
generated potential and  summarize its observational consequences.
The final section is devoted to conclusions and discussion.

%%%%%%%%%%%%%%%%%%%%%%%%%%%%%%%%%%%%%%%%%%
\section{Dynamical Generation of the Inflaton Potential}
\label{sec:potential}
Let us discuss how the fractional power-law potential
of the inflaton is generated dynamically.
First, we consider an $SP(N)$ supersymmetric gauge theory%
\footnote{We use the convention where $SP(1)$ is $SU(2)$.}
with $2(N+2)$ 
chiral superfields in the fundamental representation, $Q^I$
($I=1\cdots 2(N+2)$).
Besides the fundamental representations, we also introduce
$(N+2)(2N+3)$ gauge-singlet chiral superfields $\mathcal{Z}_{IJ} (=-\mathcal{Z}_{JI})$ which
couple to the fundamental representations in the superpotential via
\begin{eqnarray}
\label{eq:tree1}
 W = \frac{1}{2}\l_{IJ} \mathcal{Z}_{IJ}Q^IQ^J\ ,
\end{eqnarray}
with coupling constants $\l_{IJ}$.
It should be noted that all the quantum moduli, $Q^IQ^J$, are lifted 
by the couplings to the gauge singlets $\mathcal{Z}^{IJ}$.

For a later purpose, we decompose the above fields into,
\begin{eqnarray}
\begin{array}{cclccl}
    Z_{ij}  &  =   &  \mathcal{Z}_{IJ=ij}\ ,                          &    Q^i    &  =   &   Q^{I =i}\ , \\
     T_i     &  =   &  \mathcal{Z}_{IJ=i(2N+3)} \ ,      &    P       &  =   &   Q^{I = 2N+3}\ , \\
\bar{T}_i &  =  &  \mathcal{Z}_{IJ=i(2N+4)}\ , &    \bar{P} &=&  Q^{I = 2N+4}\ , \\
   S         &  =   &  \mathcal{Z}_{IJ=(2N+3)(2N+4)} \ , &  &  
\end{array}
\end{eqnarray}
where $i,j=1\cdots 2(N+1)$.
In terms of these fields, the above superpotential is now rewritten as,
\begin{eqnarray} 
\label{eq:tree2}
 W = \frac{1}{2}\l Z_{ij} Q^iQ^j + \l_T T_i Q^i P  + \l_{\bar{T}} \bar{T}_i Q^i \bar{P}
 - \l_S S P\bar{P} .
\end{eqnarray}
We have assumed $\l_{ij} = \l$ for simplicity and we also assume that $\l_{T,\bar{T}}$ 
are larger than $\l$ in the following.
The sign convention for $\l_S$ is just for later convenience.
As we will show, the scalar potential for $S$ generated by the strong dynamics features a fractional
power and eventually plays the role of the inflaton potential.

To see how the scalar potential is generated, let us  remember 
that the $SP(N)$ gauge theory with $2(N+2)$ fundamental representations 
exhibits the so-called $s$-confinement\,\cite{Seiberg:1994bz,Intriligator:1995ne}
at low energies below the dynamical scale $\Lambda$.
In this phase, the model is well described by the composite fields,
$M \propto QQ$, $M_P \propto QP$, $M_{\bar P} \propto Q \bar{P} $ and $M_{P\bar P}
\propto P\bar{P}$\,\cite{Seiberg:1994bz}, which may be assembled in the same way
in an antisymmetric matrix $\mathcal{M}$ as the gauge singlets
$Z$, $T$, $\bar{T}$, and $S$ are assembled in the antisymmetric matrix $\mathcal{Z}$.
Their effective  superpotential is given by%
\footnote{%
It should be noted that  we have neglected $O(1)$ differences  between
the $\l$'s in Eq.\,(\ref{eq:tree2}) and the ones in Eq.\,(\ref{eq:sc}) 
due to non-perturbative effects.
We also assume that the composite fields in $\mathcal{M}$ are close to the canonically normalized ones.}
\begin{eqnarray}
\label{eq:sc}
&& \!\!\!\!\!\!W_{\rm eff} = \frac{ Pf^{(N+2)}(\mathcal{M})}{\L^{(N+2)-3}} \\
+&&\!\!\!\!\!\!
\frac{1}{2}\l \L Z_{ij}M^{ij}  + \l_T \L T_i M_P^i + \l_{\bar{T}}\L \bar{T}_i M_{\bar P}^i
 - \l_S  \L S M_{P\bar{P}} .
   \nonumber 
\end{eqnarray}
The first term is the non-perturbative potential generated by the $s$-confinement.%
\footnote{%
In this paper, we define the Pfaffian of a $2n\times 2n$ antisymmetric matrix, 
$Pf^{(n)}$, so that the symplectic form $J$,
where $J = \1_n\otimes i\s_2$ with $\1_n$ being the $n\times n$ 
unit matrix and $\s_2$ the second Pauli matrix, satisfies $Pf^{(n)}(J)=1$.}
The other four terms in Eq.\,(\ref{eq:sc}) can be regarded as mass-mixing operators
between the composite fields and the gauge singlets, inducing supersymmetric masses
of $\mathcal{O}(\lambda\Lambda,\lambda_T\Lambda,\lambda_{\bar{T}}\Lambda,\lambda_S\Lambda)$,
respectively.
The effective  superpotential shows that the model possesses a
supersymmetric vacuum in which all of the $M$'s and singlets vanish.
That is, as expected, there is a vacuum with vanishing vacuum energy.  

Now, let us consider an effective potential for $S\neq 0$ and  $X (\propto J^{ij} Z_{ij})\neq 0$
around the vacuum.
Notice that  the mesons $M_{P}$ and $M_{\bar P}$  are still fixed at 
\begin{eqnarray}
 M_{P} = 0,  \quad M_{\bar{P}} = 0 \ ,
\end{eqnarray}
since we have assumed $\l_{T,\bar T}\gg \l$.
Thus, the effective potential in Eq.\,(\ref{eq:sc}) is reduced to
\begin{eqnarray}
\label{eq:sc2}
 W_{\rm eff} &=& M_{P \bar P}
\left( \frac{ Pf^{(N+1)}(M)}{\L^{(N+2)-3}}
- \l_S \L S \right)
+\frac{1}{2} \l \L Z_{ij}M^{ij}  \ .
   \nonumber 
\end{eqnarray}
The first term leads to the so-called deformed moduli constraint 
on the moduli space of the $SP(N)$ gauge theory with $2(N+1)$ fundamental
representations\,\cite{Seiberg:1994bz}.
Therefore, for a given non-vanishing $S$,
the above model is nothing but the dynamical supersymmetry breaking 
model of Ref.\,\cite{Izawa:1996pk}.
By solving the quantum deformed constraint for $S\neq 0$, we obtain
\begin{eqnarray}
\label{eq:Mij}
 M_{ij} = \Lambda \left(\frac{\l_SS}{\Lambda}\right)^{\frac{1}{N+1}}\times J_{ij}\ ,
\end{eqnarray}
which leads to 
\begin{eqnarray}
\label{eq:sc3}
 W_{\rm eff} &=&  \l \, (N+1)^{1/2} \L^2  \left(\frac{\l_SS}{\Lambda}\right)^{\frac{1}{N+1}} X  \ ,
\end{eqnarray}
where $X$ is defined by $X = Z_{ij}J^{ij}/(2(N+1)^{1/2})$.
Hence, for $S\neq 0$, supersymmetry is broken 
by the $F$-component of $X$, which leads to a scalar potential for $S$,
\begin{eqnarray}
  V \simeq \l^2 (N+1)\L^4    \left(\frac{\l_S\left|S\right|}{\Lambda}\right)^{\frac{2}{N+1}}\ .
  \label{eq:Vfracpower}
\end{eqnarray}
As promised, we find that the scalar component of the singlet $S$
obtains a fractional power-law potential,
\begin{eqnarray}
 V \propto \left|S\right|^{p}\ .
\end{eqnarray}
Its power is solely determined by the size of the $SP(N)$ gauge group,
\begin{eqnarray}
 p = \frac{2}{N+1}\ .
\end{eqnarray} 

In the above analysis, we have tacitly assumed that the field value of $S$ is around or 
below the dynamical scale, i.e.\ $\l_S S \lesssim \L$.
The above scalar potential is, however, also obtained 
for $\l_S S \gg \L$.
For that purpose, let us remember that $P$ and $\bar{P}$ are 
heavier than the dynamical scale for $\l_S S\gg \L$ 
and decouple perturbatively.
Thus, the effective theory below $\l_S S$ consists of the $SP(N)$ gauge theory 
with $2(N+1)$ fundamental representations 
whose effective dynamical scale is given by\,\cite{Shifman:1986zi}
\begin{eqnarray}
 \L_{\rm eff} = \L \times\left( \frac{\l_S\left|S\right|}{\L} \right)^{\frac{1}{2(N+1)}}\ .
\end{eqnarray}
Then, since the effective theory below $\l_S S$ is again the dynamical supersymmetry 
breaking model, we again reach the effective superpotential in Eq.\,(\ref{eq:sc3}).
Therefore, again, supersymmetry is broken by the $F$-component of $X$
at $S\neq 0$, which again leads to%
\footnote{%
The agreement between the powers in the potential in the two different field regimes
is not a coincidence but can be understood by
remembering that the effective superpotential 
consistent with (anomalous) $R$-symmetries, holomorphicity
and dimensional analysis should have the form in Eq.\,(\ref{eq:sc3}), regardless
of the sizes of $X\neq 0$ and $S\neq 0$.
}
\begin{eqnarray}
\label{eq:spot2}
  V \simeq \l^2 (N+1)\L^4    \left(\frac{\l_S\left|S\right|}{\Lambda}\right)^{\frac{2}{N+1}}\ .
\end{eqnarray}
In the following sections, we assume that $S$ 
plays the role of the inflaton.
After inflation, $S$ reaches its origin, which leads to the restoration of supersymmetry.

%%%%%%%%%%%%%%%%%%%%%%%%%%%%%%%%%%%%%%%%%%
\section{Supergravity Contributions}
\label{sec:sugra}
In our model, we apply the above obtained fractional potential to 
the chaotic inflation scenario, where 
the field value of the inflaton exceeds the Planck scale $M_{\rm Pl}$.
For such a large field value, 
we need to carefully examine the supergravity contributions to the scalar
potential, which could change the potential drastically from the fractional power-law
potential.
In fact, if we assume, for example, a minimal K\"ahler potential $S^\dagger S$ for the inflaton $S$,
the inflaton potential is modified in supergravity as follows,
\begin{eqnarray}
 V \simeq e^{\left|S\right|^2/M_{\rm Pl}^2}\times  \l^2 (N+1)\L^4 
 \left(\frac{\l_S\left|S\right|}{\Lambda}\right)^{\frac{2}{N+1}}\ ,
\end{eqnarray}
which is too steep for chaotic inflation for $S \gg M_{\rm Pl}$.%
\footnote{Even for $S<M_{\rm Pl}$, there is an eta problem. This problem
is also avoided by the solution mentioned below.}

The most attractive way to avoid such supergravity contributions is to introduce 
a shift symmetry in the direction of $S$\,\cite{Kawasaki:2000yn,Kallosh:2011qk}, 
\begin{eqnarray}
\label{eq:shift}
 S \to S + i c\ ,\quad c\in {\mathbb R}
\end{eqnarray}
(see also Ref.\,\cite{Kallosh:2010ug} for recent developments). 
With this shift symmetry, the K\"ahler potential is a function of $S+S^\dagger$, 
\begin{eqnarray}
\label{eq:Kshif}
 K  = \frac{1}{2}\left| S + S^\dagger \right|^2 + \cdots\ ,
\end{eqnarray}
and hence, it does not depend on the imaginary component of $S$, $\Im(S)$.
Therefore, the imaginary component of $S$ has a fractional power-law potential even
for $\Im(S) \gg M_{\rm Pl}$, while the real part of $S$ obtains
a Hubble mass term around $\Re(S) = 0$.
In the following,  $\sqrt{2}\Im(S)$ plays the role of the inflaton
in the chaotic inflation scenario with the dynamically generated fractional power-law potential 
in Eq.\,(\ref{eq:spot2}).%
\footnote{%
One may assume $R$-symmetry with the following charge assignments:
$Q(0)$, $P(1)$, $\bar{P}(1)$, $Z(2)$, $T(1)$, $\bar{T}(1)$,
and $S(0)$. 
With these charge assignments,
$R$-symmetry allows the shift symmetry only for the singlet $S$,
which explains why only the imaginary part of $S$ 
can exceed the Planck scale.}

One caveat to the above argument is that the shift symmetry is explicitly broken by
the interaction
\begin{eqnarray}
W \supset - \l_S S P\bar{P} 
\end{eqnarray}
in Eq.\,(\ref{eq:tree2}).
With this explicit breaking, the K\"ahler potential in Eq.\,(\ref{eq:Kshif})
obtains a radiative correction which breaks the symmetry,
\begin{eqnarray}
\delta K \sim \frac{2N\l_S^2}{16\pi^2} |S|^2 \log \left(\frac{\mu^2}{M_{\rm Pl}^2}\right)\ ,
\end{eqnarray}
where $\m$ is a renormalization scale.%
\footnote{Here, we have assumed that the K\"ahler potential in Eq.\,(\ref{eq:Kshif})
with the shift symmetry is defined around the Planck scale.
}
This breaking term causes a steep exponential potential  
of $\Im(S)$ for $\Im(S)\gg M_{\rm Pl}$ unless $\l_S$ is small enough.
Thus, to avoid a too large breaking of the shift symmetry,
we  assume in the following analysis that $\l_S$ is  rather
suppressed.\footnote{A second reason why we have to assume $\lambda_S$ to be small
is the fact that the masses of the $P$ and $\bar{P}$ quarks, $m_P = m_{\bar{P}} \simeq \lambda_S S$,
must be smaller than the Planck scale at all times, even though $S$ might take huge values,
$S \sim 10\cdots100\,M_{\textrm{Pl}}$, during inflation.
That is, too large $\lambda_S$ would entail $m_P,m_{\bar{P}} \gtrsim M_{\textrm{Pl}}$ at some
point during inflation, causing our approach based on ordinary quantum field theory to break down.}

Furthermore, we note that in general the mere introduction of a shift symmetry
in the direction of the inflaton field $S$ does not suffice to protect chaotic inflation
from receiving disastrously large supergravity corrections.
In addition we have to require that the superpotential be of the form
$W = X f\left(S\right)$\,\cite{Kawasaki:2000yn,Kallosh:2010ug},
where $f$ is an arbitrary holomorphic function of $S$
and $X$ is a gauge singlet that can be identified as the
goldstino superfield responsible for the spontaneous breaking of supersymmetry
during inflation\,\cite{Kallosh:2010xz}.
Evidently, the effective superpotential in Eq.\,(\ref{eq:sc3}), which we equally obtained in
the small-$S$ as well as in the large-$S$ regime, is just of the
required form, with $X \propto J^{ij}Z_{ij}$ playing the role of the goldstino field.
That is why, after supplementing our model with a shift symmetry in the direction of the inflaton
field $S$, all necessary conditions for the successful implementation of chaotic inflation into
supergravity are satisfied.

%%%%%%%%%%%%%%%%%%%%%%%%%%%%%%%%%%%%%%%%%%
\section{Chaotic Inflation with a Fractional Power-Law Potential}
As we have shown, simple strongly coupled gauge dynamics are able to
generate an inflationary potential featuring a fractional power.
We have also discussed the shift symmetry of the model
which suppresses the distortions of the inflaton potential
due to the supergravity contributions.
Let us now outline the phenomenology of the model, summarize
its predictions for the inflationary observables encoded in
the CMB power spectrum and, in relation to that, discuss its testability.

Inflation starts out at an arbitrary initial value of the inflaton field
above the Planck mass $S \gg M_{\textrm{Pl}}$.
At its early stages, i.e.\ as long as $\lambda_S S \gg \Lambda$,
the $SP(N)$ gauge interactions are in the perturbative regime and 
inflation is characterized by the slow-roll motion of the inflaton in the
effective potential in Eq.\,(\ref{eq:spot2}).
Similarly, we know that at small field values, i.e.\ when $\lambda_S S \ll \Lambda$,
the system is in the $s$-confinement phase, in which $S$ and the composite mesons
have masses of $\mathcal{O}\left(\lambda_{IJ}\Lambda\right)$.
In this case the inflaton potential is given by Eq.\,(\ref{eq:Vfracpower}).
In the intermediate regime, where $\lambda_S S \simeq \Lambda$, we however lack
the ability to precisely calculate the inflaton potential, which is why we do not exactly
know how the transition from the large-$S$ to the small-$S$ regime takes place.
For instance, it might be that towards the end of inflation $S$ becomes trapped
in a metastable vacuum at a field value around $\Lambda / \lambda_S$ such it actually
never reaches the small-$S$ regime.
Assuming that the effective inflaton potential exhibits no such peculiar features
around $\Lambda / \lambda_S$, we are led to the conclusion that inflation
continues without any hindrance until the slow-roll conditions become violated at
small values of the inflaton field.

The end of inflation marks the onset of preheating,
which proceeds in a rather unconventional way in our scenario
due to the negative curvature of the inflaton potential.
In fact, so far only small-field and hybrid models of inflation
have been studied in connection with a negatively curved scalar potential,
where it was found that preheating occurs via \textit{tachyonic oscillations}\,\cite{Brax:2010ai}
of the inflaton field or \textit{tachyonic preheating}\,\cite{Felder:2000hj}, respectively.
As for large-field, i.e.\ chaotic inflation only the case of a positively
curved inflaton potential, for which preheating occurs via
\textit{parametric resonance}\,\cite{Kofman:1994rk}, has been considered up to now.
We presume that in our large-field model featuring a negatively curved
inflaton potential preheating ends up being a combination of both,
tachyonic inflaton oscillations as well as parametric resonance.
The verification of this conjecture certainly requires a more comprehensive
and ultimately numerical study.

After inflation the inflaton decays through its coupling to the Higgs fields $H_u$
and $H_d$ of the supersymmetric standard model in the K\"ahler potential,
$K \supset (S+S^\dagger)H_uH_d$, at a rate
\begin{align}
\Gamma_S \sim \frac{M_S^3}{M_{\textrm{Pl}}^2},
\label{eq:GammaS}
\end{align}
with $M_S \simeq \lambda_S\Lambda$ denoting the inflaton mass.\footnote{The
same coupling leads to the non-adiabatic production of radiation during
preheating.
As it is strongly suppressed, we assume that during preheating most of the initial
vacuum energy is transferred into non-relativistic inflaton particles and only a small
fraction into radiation.
This implies in particular that the standard definition of the reheating temperature is applicable.}
Hence, the inflaton eventually reaches the supersymmetric vacuum, in which
$S = \mathcal{M} = 0$.
The rate of the perturbative inflaton decays
directly determines the reheating temperature,
$T_{\rm R} \sim \sqrt{\Gamma_S M_{\rm Pl}}$, or,
using Eq.\,(\ref{eq:GammaS}),
\begin{align}
T_{\rm R} \sim 10^7 \,\textrm{GeV}
\left(\frac{\lambda_S}{10^{-4}}\right)^{3/2} \left(\frac{\Lambda}{10^{15}\,\textrm{GeV}}\right)^{3/2} \ .
\end{align}

It is interesting to note that non-perturbative effects, i.e.\
the formation and evaporation of Q-balls near the end of inflation\,\cite{Coleman:1985ki},
could speed up the decay of the inflaton field, thus leading to a
reheating temperature much higher than in the mere perturbative picture.
In principle, the formation of Q-balls is feasible in our model
since our effective inflaton potential is shallower than a quadratic one.
Nonetheless, we suppose that no Q-balls emerge towards the end of inflation
because, owing to its
Hubble induced mass term, the real part of $S$, $\Re(S)$, is stabilized at zero.
This presumably renders it impossible to induce inspiraling orbits in $S$ field space
continuously connected to the inflationary trajectory, which would be a necessary
prerequisite for Q-balls to occur\,\cite{Kasuya:1999wu}.
A further study of this issue is beyond the scope of
this paper and shall be carried out elsewhere.

Finally, let us summarize the implications of our
fractional power-law inflaton potential for the CMB observables
and discuss the testability of our model.
Given a potential $V(S)\propto \Lambda^4 \left(\left|S\right|/\Lambda\right)^p$,
with $p=2/(N+1)$, one finds for the power spectrum $\cal{P}_{\zeta}$ of the
curvature perturbations $\zeta$\,\cite{Lyth:1998xn}
\begin{eqnarray}
{\cal P}_{\zeta}= \frac{1}{12\pi^2p^2}
\left(\frac{\Lambda}{M_{\rm Pl}}\right)^{4-p}\left(2pN_e\right)^{1+p/2},
\end{eqnarray}
where $N_e$ is the number of e-foldings.
The observational result ${\cal P}_{\zeta}=2.42\times 10^{-9}$\,\cite{Komatsu:2010fb}
then requires the dynamical scale $\Lambda$ to be shortly below the GUT scale,
\begin{eqnarray}
\Lambda \simeq 10^{15}\,\textrm{GeV} \ .
\end{eqnarray}
The spectral index $n_s$ and  the tensor-to-scalar ratio $r$ of the
curvature perturbations are respectively given by 
\begin{eqnarray}
 n_s = 1-\frac{p+2}{2N_e}, \quad r = \frac{4p}{N_e} \ .
\end{eqnarray}
For $N_e=50$ and $N \geq 1 $, such that $0 < p \leq 1$, we obtain
$n_s = 0.97\cdots0.98$ and $r = 0.16/(N+1)$, which is consistent with the
recent CMB observations\,\cite{Komatsu:2010fb}.
The Planck experiment is expected to detect the presence of tensor
modes if $r>0.05$\,\cite{Planck:2006aa}, that is, if the inflaton potential
is generated by the strong dynamics of an $SP(1) \cong SU(2)$ gauge theory.
Future experiment such as CMBPol \cite{Baumann:2008aq}
and LiteBIRD \cite{LiteBIRD}, which are expected to reach sensitivities to
$r$ of ${\cal O}(10^{-3})$, will detect tensor modes unless the underlying gauge
group is very large, $N>{\cal O}(100)$. 

%%%%%%%%%%%%%%%%%%%%%%%%%%%%%%%%%%%%%%%%%%
\section{Conclusions and Discussion}
In this paper, we have shown that fractional power-law chaotic inflation models can be realized 
in simple supersymmetric gauge theories.
In this class of  models, the energy scale during inflation is dynamically generated by 
the dimensional transmutation due to the strong gauge dynamics.
Therefore, the models not only explain the origin of the fractional power 
in the inflationary potential but also provide a reason why the energy scale of inflation
is much smaller than the Planck scale.
We also discussed how well the model fits together with the current data on the
inflationary observables.

Several comments are in order.
In our analysis, we have confined ourselves to models with $2(N+2)$ fundamental representations.
One of the reasons for this assumption is that for models with more fundamental representations
the system is in the so-called conformal window\,\cite{Banks:1981nn}, and hence, exhibits
conformal symmetry after inflation.
In such cases, the inflaton becomes an unparticle\,\cite{Georgi:2007ek} after inflation, which may change 
the evolution of the universe drastically compared to conventional inflaton scenarios. 
Although such a possibility is intriguing,
we do not pursue it further since it is  beyond the scope of this paper.

In our scenario, we have considered models in which supersymmetry
breaking in the inflaton sector vanishes after inflation.
It is,  however, possible that some portion of supersymmetry breaking in 
the inflaton sector remains non-vanishing even after inflation,  providing the dominant
source of supersymmetry breaking in the true vacuum.
In this case, we can consolidate two well motivated new physics,
supersymmetry breaking and inflation into one model.
We will discuss this possibility elsewhere.

%\begin{thebibliography}{99}
%\end{thebibliography}


\begin{thebibliography}{99}

%\cite{Guth:1980zm}
\bibitem{Guth:1980zm} 
  A.~H.~Guth,
%  ``The Inflationary Universe: A Possible Solution to the Horizon and Flatness Problems,''
  Phys.\ Rev.\ D {\bf 23}, 347 (1981);
  %%CITATION = PHRVA,D23,347;%%
%
%\cite{Linde:1981mu}
%\bibitem{Linde:1981mu} 
  A.~D.~Linde,
%  ``A New Inflationary Universe Scenario: A Possible Solution of the Horizon, Flatness, Homogeneity, Isotropy and Primordial Monopole Problems,''
  Phys.\ Lett.\ B {\bf 108}, 389 (1982);
  %%CITATION = PHLTA,B108,389;%%
%  
%\cite{Albrecht:1982wi}
%\bibitem{Albrecht:1982wi} 
  A.~Albrecht and P.~J.~Steinhardt,
%  ``Cosmology for Grand Unified Theories with Radiatively Induced Symmetry Breaking,''
  Phys.\ Rev.\ Lett.\  {\bf 48}, 1220 (1982).
  %%CITATION = PRLTA,48,1220;%%

%\cite{Mukhanov:1981xt}
\bibitem{Mukhanov:1981xt} 
  V.~F.~Mukhanov and G.~V.~Chibisov,
%  ``Quantum Fluctuation and Nonsingular Universe. (In Russian),''
  JETP Lett.\  {\bf 33}, 532 (1981)
  [Pisma Zh.\ Eksp.\ Teor.\ Fiz.\  {\bf 33}, 549 (1981)];
  %%CITATION = JTPLA,33,532;%%
%
%\cite{Linde:2005ht}
%\bibitem{Linde:2005ht} 
  A.~D.~Linde,
%  ``Particle physics and inflationary cosmology,''
  Contemp.\ Concepts Phys.\  {\bf 5}, 1 (1990)
  [hep-th/0503203].
  %%CITATION = HEP-TH/0503203;%%

%\cite{Lyth:1998xn}
\bibitem{Lyth:1998xn} 
  D.~H.~Lyth and A.~Riotto,
%  ``Particle physics models of inflation and the cosmological density perturbation,''
  Phys.\ Rept.\  {\bf 314}, 1 (1999)
  [hep-ph/9807278].
  %%CITATION = HEP-PH/9807278;%%

%\cite{Linde:1983gd}
\bibitem{Linde:1983gd} 
  A.~D.~Linde,
%  ``Chaotic Inflation,''
  Phys.\ Lett.\ B {\bf 129}, 177 (1983);
  %%CITATION = PHLTA,B129,177;%%
%
%\cite{Linde:1984st}
%\bibitem{Linde:1984st} 
  A.~D.~Linde,
  %``Chaotic Inflating Universe,''
  JETP Lett.\  {\bf 38}, 176 (1983)
  [Pisma Zh.\ Eksp.\ Teor.\ Fiz.\  {\bf 38}, 149 (1983)].
  %%CITATION = JTPLA,38,176;%%

%%\cite{Komatsu:2008hk}
%\bibitem{Komatsu:2008hk} 
%  E.~Komatsu {\it et al.}  [WMAP Collaboration],
%  %``Five-Year Wilkinson Microwave Anisotropy Probe (WMAP) Observations: Cosmological Interpretation,''
%  Astrophys.\ J.\ Suppl.\  {\bf 180}, 330 (2009)
%  [arXiv:0803.0547 [astro-ph]];
%  %%CITATION = ARXIV:0803.0547;%%

%\cite{Starobinsky:1985ww}
\bibitem{Starobinsky:1985ww} 
  A.~A.~Starobinsky,
  %``Cosmic Background Anisotropy Induced by Isotropic Flat-Spectrum Gravitational-Wave Perturbations,''
  Sov.\ Astron.\ Lett.\  {\bf 11}, 133 (1985);
  %%CITATION = SALED,11,133;%%
%
%\cite{Lyth:1996im}
%\bibitem{Lyth:1996im} 
  D.~H.~Lyth,
  %``What would we learn by detecting a gravitational wave signal in the cosmic microwave background anisotropy?,''
  Phys.\ Rev.\ Lett.\  {\bf 78}, 1861 (1997)
  [hep-ph/9606387].
  %%CITATION = HEP-PH/9606387;%%

%\cite{Komatsu:2010fb}
\bibitem{Komatsu:2010fb} 
  E.~Komatsu {\it et al.}  [WMAP Collaboration],
%  ``Seven-Year Wilkinson Microwave Anisotropy Probe (WMAP) Observations: Cosmological Interpretation,''
  Astrophys.\ J.\ Suppl.\  {\bf 192}, 18 (2011)
  [arXiv:1001.4538 [astro-ph.CO]].
  %%CITATION = ARXIV:1001.4538;%%

%\cite{Planck:2006aa}
\bibitem{Planck:2006aa} 
  [Planck Collaboration],
%  ``The Scientific programme of planck,''
  astro-ph/0604069.
  %%CITATION = ASTRO-PH/0604069;%%

%\cite{Alabidi:2010sf}
\bibitem{Alabidi:2010sf} 
  L.~Alabidi and I.~Huston,
%  ``An update on single field models of inflation in light of WMAP7,''
  JCAP {\bf 1008}, 037 (2010)
  [arXiv:1004.4794 [astro-ph.CO]];
  %%CITATION = ARXIV:1004.4794;%%
%  
%\cite{Martin:2010hh}
%\bibitem{Martin:2010hh} 
  J.~Martin, C.~Ringeval and R.~Trotta,
%  ``Hunting Down the Best Model of Inflation with Bayesian Evidence,''
  Phys.\ Rev.\ D {\bf 83}, 063524 (2011)
  [arXiv:1009.4157 [astro-ph.CO]].
  %%CITATION = ARXIV:1009.4157;%%

%\cite{Kallosh:2010ug}
\bibitem{Kallosh:2010ug} 
  R.~Kallosh and A.~Linde,
%  ``New models of chaotic inflation in supergravity,''
  JCAP {\bf 1011}, 011 (2010)
  [arXiv:1008.3375 [hep-th]].
  %%CITATION = ARXIV:1008.3375;%%

%\cite{Kallosh:2010xz}
\bibitem{Kallosh:2010xz} 
  R.~Kallosh, A.~Linde and T.~Rube,
%  ``General inflaton potentials in supergravity,''
  Phys.\ Rev.\ D {\bf 83}, 043507 (2011)
  [arXiv:1011.5945 [hep-th]].
  %%CITATION = ARXIV:1011.5945;%%

%\cite{Takahashi:2010ky}
\bibitem{Takahashi:2010ky} 
  F.~Takahashi,
%  ``Linear Inflation from Running Kinetic Term in Supergravity,''
  Phys.\ Lett.\ B {\bf 693}, 140 (2010)
  [arXiv:1006.2801 [hep-ph]].
  %%CITATION = ARXIV:1006.2801;%%

%\cite{Silverstein:2008sg}
\bibitem{Silverstein:2008sg} 
  E.~Silverstein and A.~Westphal,
%  ``Monodromy in the CMB: Gravity Waves and String Inflation,''
  Phys.\ Rev.\ D {\bf 78}, 106003 (2008)
  [arXiv:0803.3085 [hep-th]].
  %%CITATION = ARXIV:0803.3085;%%

%\cite{Dimopoulos:1997fv}
\bibitem{Dimopoulos:1997fv} 
  S.~Dimopoulos, G.~R.~Dvali and R.~Rattazzi,
%  ``Dynamical inflation and unification scale on quantum moduli spaces,''
  Phys.\ Lett.\ B {\bf 410}, 119 (1997)
  [hep-ph/9705348];
  %%CITATION = HEP-PH/9705348;%%
%
%\cite{Izawa:1997df}
%\bibitem{Izawa:1997df} 
  K.~I.~Izawa, M.~Kawasaki and T.~Yanagida,
%  ``Dynamical tuning of the initial condition for new inflation in supergravity,''
  Phys.\ Lett.\ B {\bf 411}, 249 (1997)
  [hep-ph/9707201];
  %%CITATION = HEP-PH/9707201;%%
%
%\cite{Izawa:1997jc}
%\bibitem{Izawa:1997jc} 
  K.~I.~Izawa,
%  ``Supersymmetry-breaking models of inflation,''
  Prog.\ Theor.\ Phys.\  {\bf 99}, 157 (1998)
  [hep-ph/9708315].
  %%CITATION = HEP-PH/9708315;%%

%\cite{Seiberg:1994bz}
\bibitem{Seiberg:1994bz} 
  N.~Seiberg,
%  ``Exact results on the space of vacua of four-dimensional SUSY gauge theories,''
  Phys.\ Rev.\ D {\bf 49}, 6857 (1994)
  [hep-th/9402044];
  %%CITATION = HEP-TH/9402044;%%

%\cite{Intriligator:1995ne}
\bibitem{Intriligator:1995ne} 
  K.~A.~Intriligator and P.~Pouliot,
%  ``Exact superpotentials, quantum vacua and duality in supersymmetric SP(N(c)) gauge theories,''
  Phys.\ Lett.\ B {\bf 353}, 471 (1995)
  [hep-th/9505006];
  %%CITATION = HEP-TH/9505006;%%
%
%\cite{Csaki:1996sm}
%\bibitem{Csaki:1996sm} 
  C.~Csaki, M.~Schmaltz and W.~Skiba,
%  ``A Systematic approach to confinement in N=1 supersymmetric gauge theories,''
  Phys.\ Rev.\ Lett.\  {\bf 78}, 799 (1997)
  [hep-th/9610139].
  %%CITATION = HEP-TH/9610139;%%

%\cite{Izawa:1996pk}
\bibitem{Izawa:1996pk} 
  K.~-I.~Izawa and T.~Yanagida,
  %``Dynamical supersymmetry breaking in vector - like gauge theories,''
  Prog.\ Theor.\ Phys.\  {\bf 95}, 829 (1996)
  [hep-th/9602180];
  %%CITATION = HEP-TH/9602180;%%
%\cite{Intriligator:1996pu}
%\bibitem{Intriligator:1996pu} 
  K.~A.~Intriligator and S.~D.~Thomas,
  %``Dynamical supersymmetry breaking on quantum moduli spaces,''
  Nucl.\ Phys.\ B {\bf 473}, 121 (1996)
  [hep-th/9603158].
  %%CITATION = HEP-TH/9603158;%%

%\cite{Shifman:1986zi}
\bibitem{Shifman:1986zi} 
  M.~A.~Shifman and A.~I.~Vainshtein,
  %``Solution of the Anomaly Puzzle in SUSY Gauge Theories and the Wilson Operator Expansion,''
  Nucl.\ Phys.\ B {\bf 277}, 456 (1986)
  [Sov.\ Phys.\ JETP {\bf 64}, 428 (1986)]
  [Zh.\ Eksp.\ Teor.\ Fiz.\  {\bf 91}, 723 (1986)];
  %%CITATION = NUPHA,B277,456;%%
%
%\cite{Veneziano:1982ah}
%\bibitem{Veneziano:1982ah} 
  G.~Veneziano and S.~Yankielowicz,
  %``An Effective Lagrangian for the Pure N=1 Supersymmetric Yang-Mills Theory,''
  Phys.\ Lett.\ B {\bf 113}, 231 (1982);
  %%CITATION = PHLTA,B113,231;%%
%
%\cite{Taylor:1982bp}
%\bibitem{Taylor:1982bp} 
  T.~R.~Taylor, G.~Veneziano and S.~Yankielowicz,
  %``Supersymmetric QCD and Its Massless Limit: An Effective Lagrangian Analysis,''
  Nucl.\ Phys.\ B {\bf 218}, 493 (1983);
  %%CITATION = NUPHA,B218,493;%%
%
%\cite{Novikov:1983uc}
%\bibitem{Novikov:1983uc} 
  V.~A.~Novikov, M.~A.~Shifman, A.~I.~Vainshtein and V.~I.~Zakharov,
  %``Exact Gell-Mann-Low Function of Supersymmetric Yang-Mills Theories from Instanton Calculus,''
  Nucl.\ Phys.\ B {\bf 229}, 381 (1983);
  %%CITATION = NUPHA,B229,381;%%
%
%\cite{Seiberg:1993vc}
%\bibitem{Seiberg:1993vc} 
  N.~Seiberg,
  %``Naturalness versus supersymmetric nonrenormalization theorems,''
  Phys.\ Lett.\ B {\bf 318}, 469 (1993)
  [hep-ph/9309335].
  %%CITATION = HEP-PH/9309335;%%

%\cite{Kawasaki:2000yn}
\bibitem{Kawasaki:2000yn} 
  M.~Kawasaki, M.~Yamaguchi and T.~Yanagida,
  %``Natural chaotic inflation in supergravity,''
  Phys.\ Rev.\ Lett.\  {\bf 85}, 3572 (2000)
  [hep-ph/0004243].
  %%CITATION = HEP-PH/0004243;%%

%\cite{Kallosh:2010ug}
%\bibitem{Kallosh:2010ug} 
%  R.~Kallosh and A.~Linde,
%  %``New models of chaotic inflation in supergravity,''
%  JCAP {\bf 1011}, 011 (2010)
%  [arXiv:1008.3375 [hep-th]];
%  %%CITATION = ARXIV:1008.3375;%%

%\cite{Kallosh:2011qk}
\bibitem{Kallosh:2011qk} 
  R.~Kallosh, A.~Linde, K.~A.~Olive and T.~Rube,
  %``Chaotic inflation and supersymmetry breaking,''
  Phys.\ Rev.\ D {\bf 84}, 083519 (2011)
  [arXiv:1106.6025 [hep-th]].
  %%CITATION = ARXIV:1106.6025;%%

%\cite{Brax:2010ai}
\bibitem{Brax:2010ai} 
  P.~Brax, J.~-F.~Dufaux and S.~Mariadassou,
  %``Preheating after Small-Field Inflation,''
  Phys.\ Rev.\ D {\bf 83}, 103510 (2011)
  [arXiv:1012.4656 [hep-th]].
  %%CITATION = ARXIV:1012.4656;%%

%\cite{Felder:2000hj}
\bibitem{Felder:2000hj} 
  G.~N.~Felder, J.~Garcia-Bellido, P.~B.~Greene, L.~Kofman, A.~D.~Linde and I.~Tkachev,
  %``Dynamics of symmetry breaking and tachyonic preheating,''
  Phys.\ Rev.\ Lett.\  {\bf 87}, 011601 (2001)
  [hep-ph/0012142];
  %%CITATION = HEP-PH/0012142;%%
%
%\cite{Felder:2001kt}
%\bibitem{Felder:2001kt} 
  G.~N.~Felder, L.~Kofman and A.~D.~Linde,
  %``Tachyonic instability and dynamics of spontaneous symmetry breaking,''
  Phys.\ Rev.\ D {\bf 64}, 123517 (2001)
  [hep-th/0106179].
  %%CITATION = HEP-TH/0106179;%%

%\cite{Kofman:1994rk}
\bibitem{Kofman:1994rk} 
  L.~Kofman, A.~D.~Linde and A.~A.~Starobinsky,
  %``Reheating after inflation,''
  Phys.\ Rev.\ Lett.\  {\bf 73}, 3195 (1994)
  [hep-th/9405187];
  %%CITATION = HEP-TH/9405187;%%
%
%\cite{Kofman:1997yn}
%\bibitem{Kofman:1997yn} 
  L.~Kofman, A.~D.~Linde and A.~A.~Starobinsky,
  %``Towards the theory of reheating after inflation,''
  Phys.\ Rev.\ D {\bf 56}, 3258 (1997)
  [hep-ph/9704452].
  %%CITATION = HEP-PH/9704452;%%

%\cite{Coleman:1985ki}
\bibitem{Coleman:1985ki} 
  S.~R.~Coleman,
  %``Q Balls,''
  Nucl.\ Phys.\ B {\bf 262}, 263 (1985)
  [Erratum-ibid.\ B {\bf 269}, 744 (1986)];
  %%CITATION = NUPHA,B262,263;%%
%
%\cite{Cohen:1986ct}
%\bibitem{Cohen:1986ct} 
  A.~G.~Cohen, S.~R.~Coleman, H.~Georgi and A.~Manohar,
  %``The Evaporation Of Q Balls,''
  Nucl.\ Phys.\ B {\bf 272}, 301 (1986);
  %%CITATION = NUPHA,B272,301;%%
%
%\cite{Enqvist:2002si}
%\bibitem{Enqvist:2002si} 
  K.~Enqvist, S.~Kasuya and A.~Mazumdar,
  %``Inflatonic solitons in running mass inflation,''
  Phys.\ Rev.\ D {\bf 66}, 043505 (2002)
  [hep-ph/0206272].
  %%CITATION = HEP-PH/0206272;%%

%\cite{Kasuya:1999wu}
\bibitem{Kasuya:1999wu} 
  S.~Kasuya and M.~Kawasaki,
  %``Q ball formation through Affleck-Dine mechanism,''
  Phys.\ Rev.\ D {\bf 61}, 041301 (2000)
  [hep-ph/9909509];
  %%CITATION = HEP-PH/9909509;%%
%
%\cite{Jokinen:2002xw}
%\bibitem{Jokinen:2002xw} 
  A.~Jokinen,
  %``Analytical and numerical properties of Affleck-Dine condensate formation,''
  hep-ph/0204086.
  %%CITATION = HEP-PH/0204086;%%

%\cite{Baumann:2008aq}
\bibitem{Baumann:2008aq} 
  D.~Baumann {\it et al.}  [CMBPol Study Team Collaboration],
  %``CMBPol Mission Concept Study: Probing Inflation with CMB Polarization,''
  AIP Conf.\ Proc.\  {\bf 1141}, 10 (2009)
  [arXiv:0811.3919 [astro-ph]].
  %%CITATION = ARXIV:0811.3919;%%

\bibitem{LiteBIRD}
LiteBIRD project,

\verb$http://cmb.kek.jp/litebird/index.html$.

%\cite{Banks:1981nn}
\bibitem{Banks:1981nn} 
  T.~Banks and A.~Zaks,
  %``On the Phase Structure of Vector-Like Gauge Theories with Massless Fermions,''
  Nucl.\ Phys.\ B {\bf 196}, 189 (1982).
  %%CITATION = NUPHA,B196,189;%%

%\cite{Georgi:2007ek}
\bibitem{Georgi:2007ek} 
  H.~Georgi,
  %``Unparticle physics,''
  Phys.\ Rev.\ Lett.\  {\bf 98}, 221601 (2007)
  [hep-ph/0703260].
  %%CITATION = HEP-PH/0703260;%%

\end{thebibliography}
\end{document}